\documentclass[12pt]{iopart}

\usepackage{amsmath}
\usepackage{graphicx}
\usepackage{bm}
\usepackage[dvipsnames]{xcolor}
\usepackage{epstopdf}
\usepackage{float}
\usepackage{cite}
\usepackage{multirow}
\usepackage{appendix}
\usepackage{balance}
\usepackage{silence}
\newcommand{\bra}[1]{\langle#1|} 
\newcommand{\ket}[1]{|#1\rangle} 
\newcommand{\braket}[2]{ \langle #1 | #2 \rangle} 
\newcommand{\nor}[1]{|#1|^2}
\usepackage{iopams}  
\begin{document}

\title[]{Quantum cloning transformation unlocks the potential of W class of states in a controlled quantum secure direct communication protocol}

\author{Rashi Jain  \& Satyabrata Adhikari$^*$}

\address{Department of Applied Mathematics,\\ Delhi Technological University, Delhi-110042, India\\
$^*$\begin{footnotesize}Author to whom any correspondence should be addressed.\end{footnotesize}}
\ead{rashijain\_23phdam07@dtu.ac.in, satyabrata@dtu.ac.in }
\vspace{10pt}
\begin{indented}
\item[]August 2017 (minor update March 2024)
\end{indented}

	
	
	
	\begin{abstract}
		In a controlled quantum secure direct communication (Controlled QSDC) protocol between three parties, the sender sends the encoded secured message to one of the two receivers, which can be decoded only when the other receiver agrees to cooperate. A lot of studies have been done on it using the three-qubit GHZ state, and only a few works have involved the W state. In this work, we introduce a controlled QSDC protocol exploiting a three-qubit W class of state shared between three parties, Alice (Sender), Bob (Controller), and Charlie (Receiver). In the proposed protocol, the shared state parameters and the secret are linked in such a way that it is very difficult to factor them. We will show that these parameters can be factored out easily if the receiver uses a quantum cloning machine (QCM) and thus can retrieve the secret. We find that the protocol is probabilistic and have calculated the probability of success of the protocol. Further, we establish the relation between the success probability and the efficiency of the QCM. In general, we find that the efficiency of the constructed QCM is greater than or equal to $\frac{1}{3}$, but we have shown that its efficiency can be enhanced when the parameters of the shared state are used as the parameters of the QCM. Moreover, we derived the linkage between the probability of success and the amount of entanglement in the shared W class of state. We analyzed the obtained result and found that even a less entangled W class of state can also play a vital role in the proposed controlled QSDC scheme.\\
		
	\end{abstract}
	
	\maketitle{}
	
		\section{Introduction}
	The transfer of confidential information plays a vital role in almost all industries today. It has now become a task of utmost importance to ensure the security of the information sent to the desired receiver when eavesdropper is present between the sender and the receiver. Therefore, it is important first to construct a secret key to transmit the information later. Quantum cryptography \cite{gisin_2002} provides a basis for doing so. A type of quantum secure direct communication protocol, namely QSDC \cite{beige_2002} has been proposed that allows secure communication among two distant parties without establishing a shared secret key to encrypt it. Controlled QSDC provides one of the most efficient schemes in which a secret information is sent to the receiver in the presence of a controller. The secret can be obtained by the receiver only when the controller agrees to cooperate with him/her. This uses the most powerful tools of quantum mechanics \cite{ballentine_2014}, such as quantum entanglement \cite{horodecki_2009}, quantum teleportation \cite{bennett_1995} and quantum cloning \cite{scarani_2005, adhikari_thesis}.\\
	Quantum cloning plays a vital role in quantum cryptography. In this regard, Wooters and Zurek \cite{wooters_1982}, in 1982, stated that "an arbitrary quantum state cannot be cloned", famously known as the no-cloning theorem. This result makes one of the major differences between classical and quantum information processing. Since perfect cloning is not possible, an approximate QCM can be constructed. The QCM proposed by Wooters and Zurek (W-Z copying machine) is given by \cite{wooters_1982}
	\begin{equation}
		\begin{split}
			\ket{0}_a\ket{Q}_x\longrightarrow \ket{0}_a\ket{0}_b\ket{Q_0}_x\\ 
			\ket{1}_a\ket{Q}_x\longrightarrow \ket{1}_a\ket{1}_b\ket{Q_1}_x
		\end{split}
	\end{equation}
	where $\ket{Q}_x$ denote the QCM state vector, subscript $a$ denotes the original input state and subscript $b$ denotes the copy mode of the input state. The copying machine state vector satisfy the following
	\begin{eqnarray}
		\braket{Q_0}{Q_1}=0,~ \braket{Q_i}{Q_i}=1,~ i=0,1
	\end{eqnarray}
	The W-Z copying machine is state-dependent \cite{bruss_1998_cloning}, i.e., the quality of the copies at the output of the quantum cloner depends on the input state. However, there are quantum cloners that are state-independent \cite{buzek_1998}. Going towards in this direction, Buzek and Hillery \cite{buzek_1996} analyzed the possibility of approximately copying an arbitrary quantum state by using a universal quantum cloning machine (UQCM). UQCM is a type of quantum cloning machine in which the quality of the copy is independent of the input state. The following transformation describes the optimal UQCM
	\begin{equation}
		\begin{split}
			\ket{0}_a\ket{0}_b\ket{Q}_x\longrightarrow \sqrt{\frac{2}{3}}\ket{00}_{ab}\ket{0}_x + \sqrt{\frac{1}{3}}\ket{\xi}_{ab}\ket{1}_x\\
			\ket{1}_a\ket{0}_b\ket{Q}_x\longrightarrow \sqrt{\frac{2}{3}}\ket{11}_{ab}\ket{1}_x + \sqrt{\frac{1}{3}} \ket{\xi}_{ab}\ket{0}_x
		\end{split}
	\end{equation}
	where $\ket{\xi}=\frac{\ket{01}+\ket{10}}{\sqrt{2}}$.\\
	Due to the property that perfect copies of an arbitrary state cannot be made, one cannot distribute the shares of a quantum secret to two or more parties simultaneously. Thus, it makes the no-cloning theorem a fundamental constraint in a controlled QSDC scheme.\\
	In a QSDC protocol, a sender distributes the secret message among all the participants, and each participant plays an equal and important role in successfully completing the protocol. This also ensures that if there is a dishonest member in the group, he may not be able to extract the secret without the cooperation of other participants. The idea of QSDC has gain a lot of attention in the recent years because of its following advantages: (i) QSDC is assumed to be secured from any classical attack and also even from quantum attack where an attacker/eavesdropper possesses powerful equipment such as quantum computer, (ii) Security of the information being transferred without a key through quantum channel is ensured by the laws of quantum mechanics in a QSDC protocol \cite{ahn_2024}, and (iii) Entangled state serves as a quantum channel in a QSDC protocol ensures the security against noise or presence of an eavesdropper \cite{qi_2021}. \\
  Exploiting the properties of entanglement, the first QSDC protocol was put forward by Long and Liu, in 2000 \cite{long_2000}. In 2002, a Bell-state based QSDC protocol, popularly known as ping-pong protocol was proposed by Bostrom and Felbinger \cite{bostrom_2002}. A two-step QSDC protocol using blocks of EPR pairs, and a protocol based on single photon four states are proposed in \cite{deng_2003 ,deng_2004}. Cai et al. \cite{cai_2004} studied a QSDC protocol only using single qubit in a mixed state. Other QSDC protocols such as the decoy-state QSDC \cite{park_2023}, device independent QSDC \cite{roy_2024}, multi-party controlled QSDC \cite{basak_2023}, bidirectional QSDC \cite{feng_2025}, and semi-QSDC protocols \cite{rong_2020} have been discussed in the recent years. Based on similar concepts, some other protocols such as semi-quantum private comparison protocol \cite{chongqiang_2021,gong_2024}, where the two parties compare the equality of their private information with the help of a semi-honest third party, and semi-quantum key distribution protocols \cite{zhou_2024}, which enables two parties to establish a secret key with the help of a third party, have also been proposed. Although most of the work on controlled QSDC schemes has been done using the GHZ states \cite{khorasani_2020}, there have been only a few with W-states \cite{chen_2008}. Recently, in 2024, Pan et al. \cite{pan_2024} studied various QSDC protocols and listed their challenges and future scope. They highlighted the recent theoretical advancements and the practical experimentation in the field.\\
In this work, we propose a controlled QSDC scheme among three parties, one sender, one controller, and one receiver. We, therefore require atleast a three-qubit shared state among them. Based on the classification of three-qubit states, we have three-qubit genuine entangled states of two types namely, the GHZ class of states and the W class of states. The major difference between the two is that, under the local operation and classical communication (LOCC), these states cannot be transformed into one another \cite{dur_2000}. In the present work, we would like to explore the pairwise entanglement property of W-class of states in a controlled QSDC protocol, as they are considered much more robust than GHZ class of states \cite{chen_2008}. The entanglement in W states is preserved even after measurements are performed on a single particle, making them suitable for quantum communications \cite{jian_2007}. The motivations of this work are as follows: (i) Because of the robustness characterstics of W class of states in terms of particle loss \cite{vijayan_2020}, we consider W class of states to study its effect in controlled QSDC protocol and, (ii) to examine how the amount of entanglement contained in the W-class of states affects the probability of success of the controlled QSDC protocol. (iii) Since our proposed controlled QSDC protocol does not require any key so the security of information is assured \cite{ying_2024}. \\
	The rest of the paper is organized as follows: In section II, we propose a symmetric QCM and study how better it can create a copy at the output of an arbitrary quantum state in the input. In section III, we propose a controlled QSDC protocol for a three-party system using a W-class of state. We, further, analyzed the proposed controlled QSDC scheme and calculated the probability of success of it. Moreover, we discussed in detail the relation between the probability of success of the controlled QSDC scheme and the efficiency of the proposed QCM. In section IV, the linkage between the amount of entanglement of the shared W class of state and the probability of success of the controlled QSDC scheme is studied. In section V, we present the comparison of our work with the existing protocol, and the pros and cons of our study. We also discuss various internal and external attacks possible on the proposed QSDC protocol in the security analysis section VI. We finally conclude in section VII.
	
	\section{Symmetric Quantum Cloning Transformation}
	In this section, we propose a symmetric QCM, which will be used in the later section, where the controlled QSDC protocol will be discussed. Let us consider an arbitrary single-qubit state $\ket{\chi}^{in}$ that we would like to clone. The input state $\ket{\chi}^{in}$ can be expressed in the form 
	\begin{eqnarray}
		\ket{\chi}^{in}=x\ket{0}+y\ket{1},~~\nor{x}+\nor{y}=1
	\end{eqnarray}
	To make a copy of the input state, we propose the quantum cloning transformation, which is of the form
	\begin{equation}
		\begin{split}
			U\ket{0}_{a}\ket{0}_{b}\ket{Q}_{c}=\big(\ket{00}_{ab}+p(\ket{01}_{ab}+\ket{10}_{ab})\big)\ket{Q_0}_{c}\\
			U\ket{1}_{a}\ket{0}_{b}\ket{Q}_{c}=\big(\ket{11}_{ab}+q(\ket{01}_{ab}+\ket{10}_{ab})\big)\ket{Q_1}_{c}
			\label{qcm_gen}
		\end{split}
	\end{equation}
	where $U$ denotes an unitary operator, $\ket{0}_{a}$, $\ket{0}_{b}$ and $\ket{Q}_{c}$ denote the input qubit mode, blank qubit mode and cloning machine state vector respectively. The cloning machine state vector at the output is denoted by $\ket{Q_0}_{c}$ and $\ket{Q_1}_{c}$ respectively. Here we note that $p$ and $q$ may represent the complex parameters. \\
	Let us make the following assumption 
	\begin{equation}
		\braket{Q_0}{Q_1}=0
	\end{equation}
	The unitarity of the transformation (\ref{qcm_gen}) gives
	\begin{equation}
		\begin{split}
			\braket{Q_0}{Q_0}=\frac{1}{1+2\nor{p}}\\
			\braket{Q_1}{Q_1}=\frac{1}{1+2\nor{q}}
			\label{unitarity_qcm_gen}
		\end{split}
	\end{equation}
	The cloning transformation (\ref{qcm_gen}), when acts on an arbitrary single qubit quantum state, we obtain
	\begin{eqnarray}
		(x\ket{0}_a+y&\ket{1}_a)\ket{0}_b\ket{Q}_c\notag\\
		=&x\ket{00}_{ab}\ket{Q_0}_c+xp(\ket{01}_{ab}+\ket{10}_{ab})\ket{Q_0}_c+\notag\\
		&y\ket{11}_{ab}\ket{Q_1}_c+yq(\ket{01}_{ab}+\ket{10}_{ab})\ket{Q_1}_c
		\label{trans_gen}
	\end{eqnarray}
	For the reduced density operator $\rho_{ab}$ after tracing out the machine state vectors, we have
	\begin{eqnarray}
		\rho_{ab}=&\nor{x}\ket{00}_{ab}\bra{00}\braket{Q_0}{Q_0}+2\nor{x}\nor{p}\ket{\xi}_{ab}\bra{\xi}\braket{Q_0}{Q_0}+\notag\\
		&\nor{y}\ket{11}_{ab}\bra{11}\braket{Q_1}{Q_1}+2\nor{y}\nor{q}\ket{\xi}_{ab}\bra{\xi}\braket{Q_1}{Q_1}
		\label{trace_q_gen}
	\end{eqnarray}
	where $\ket{\xi}=\frac{\ket{01}+\ket{10}}{\sqrt{2}}$. \\
	The reduced density operator describing the original mode can be obtained by taking a partial trace over the copy mode, which gives
	\begin{eqnarray}
		\rho_{a}=& \nor{x}\ket{0}\bra{0}\braket{Q_0}{Q_0}+2\nor{x}\nor{p}(\ket{0}\bra{0}+\ket{1}\bra{1})\braket{Q_0}{Q_0}+\notag\\
		&\nor{y}\ket{1}\bra{1}\braket{Q_1}{Q_1}+2\nor{y}\nor{q}(\ket{0}\bra{0}+\ket{1}\bra{1})\braket{Q_1}{Q_1}\notag\\
		=& \big[\nor{x}(1+2\nor{p})\braket{Q_0}{Q_0}+2\nor{y}\nor{q}\braket{Q_1}{Q_1}\big]\ket{0}\bra{0}+\notag\\
		&\big[2\nor{x}\nor{p}\braket{Q_0}{Q_0}+\nor{y}(1+2\nor{q})\braket{Q_1}{Q_1}\big]\ket{1}\bra{1}
		\label{trace_c2_gen}
	\end{eqnarray}
	Similarly, tracing out the mode $a$ from (\ref{trace_q_gen}), we get
	\begin{eqnarray}
		\rho_{b}=& \nor{x}\ket{0}\bra{0}\braket{Q_0}{Q_0}+2\nor{x}\nor{p}(\ket{1}\bra{1}+\ket{0}\bra{0})\braket{Q_0}{Q_0}+\notag\\
		&\nor{y}\ket{1}\bra{1}\braket{Q_1}{Q_1}+2\nor{y}\nor{q}(\ket{1}\bra{1}+\ket{0}\bra{0})\braket{Q_1}{Q_1}\notag\\
		=& \big[\nor{x}(1+2\nor{p})\braket{Q_0}{Q_0}+2\nor{y}\nor{q}\braket{Q_1}{Q_1}\big]\ket{0}\bra{0}+\notag\\
		&\big[2\nor{x}\nor{p}\braket{Q_0}{Q_0}+\nor{y}(1+2\nor{q})\braket{Q_1}{Q_1}\big]\ket{1}\bra{1}
		\label{trace_c1_gen}
	\end{eqnarray}
	Since it is now clear that the two output states in the original mode and copy mode are described by the density operators $\rho_{a}$ and $\rho_{b}$ are identical, so we can say that the prescribed transformation (\ref{qcm_gen}) describes a symmetric QCM.\\
	To know how good the cloning machine is, i.e., how close is the output mode from the initial mode, we study the efficiency of the prescribed QCM. The efficiency of the cloning machine can be described by the Hilbert-Schmidt (HS) distance \cite{witte_1999}, which can be defined as
	\begin{equation}
		D_i=\text{Tr}\big[(\rho^{in}-\rho_{i})(\rho^{in}-\rho_{i})^\dagger\big],~~i=a,b
		\label{d_a_formula}
	\end{equation}
	where $\rho^{in}=\ket{\chi}^{in}\bra{\chi}$ denotes the input density operator and can be expressed as
	\begin{eqnarray}
		\rho^{in}=\nor{x}\ket{0}\bra{0}+xy^*\ket{0}\bra{1}+yx^*\ket{1}\bra{0}+\nor{y}\ket{1}\bra{1}
	\end{eqnarray}
	Now we are in a position to calculate $D_a$ and $D_{b}$. Since the density operators $\rho_{a}$ and $\rho_{b}$ are equal so $D_{a}$ and $D_{b}$ will be equal. Therefore, the HS distance can be calculated as  
	\begin{eqnarray}
			D_a=\text{Tr}[(\rho^{in}-\rho_{a})(\rho^{in}-\rho_{a})^\dagger]\cr
			=2|x|^4\left(\frac{2|q|^4}{(1+\nor{q})^2}+\frac{2|p|^4}{(1+\nor{p})^2}-1\right)+\cr
			2\nor{x}\left(1-\frac{4|q|^4}{(1+\nor{q})^2}\right)+\frac{4|q|^4}{(1+\nor{q})^2}
			\label{d_a_gen}
	\end{eqnarray}
	
	It can be seen from the above expression (\ref{d_a_gen}) that we can make the proposed quantum cloning transformation state independent if we choose the parameters $\nor{p}=\nor{q}=1$. However, in this scenario, the HS distance $D_a$ will be equal to unity and may be considered very large, and thus, it will make the cloning transformation inefficient. For other values of $\nor{p}$ and $\nor{q}$, the HS distance $D_a$ will be a function of the input state parameter $\nor{x}$, and thus, the cloning transformation will be state-dependent. This indicates the fact that for some input states, the cloning transformation will be better, and for some, it will be worse. Thus, for state-dependent cloning transformation, it would be better to consider the average HS distance to determine the efficiency of the cloning transformation \cite{kumar_2020}. The average HS distance can be defined as
	\begin{eqnarray}
		\overline{D_a}=\int_{0}^{1} D_a \text{d}\nor{x}
		\label{d_a_bar_gen}
	\end{eqnarray}
	To find the efficiency of the proposed cloning transformation (\ref{qcm_gen}), let us calculate the average HS distance. To do this, let us substitute (\ref{d_a_gen}) in (\ref{d_a_bar_gen}), we get
	\begin{eqnarray}
		\overline{D_a}=
		\frac{4}{3}\bigg[\frac{|q|^4}{(1+2\nor{q})^2}+\frac{|p|^4}{(1+2\nor{p})^2}\bigg]+\frac{1}{3}
		\label{avhsdis}
	\end{eqnarray}
	We are now in a position to analyze the efficiency of the cloning transformation. Since the first term of (\ref{avhsdis}) is always non-negative so $\overline{D_a}\geq \frac{1}{3}$. The equality is achieved only when $\nor{p}=\nor{q}=0$ and in this case, the proposed cloning transformation reduces to the WZ quantum cloning transformation \cite{wooters_1982}. Thus, we may say that the prescribed quantum cloning transformation may not be an efficient quantum cloning transformation but we will show in the subsequent section that it will play a vital role in retrieving the secret in the controlled QSDC protocol.

	\section{Controlled QSDC protocol using a class of W states and quantum cloning transformation}
	
	\subsection{Proposed controlled QSDC protocol}
	In this section, we introduce a controlled QSDC protocol using a W-class of state shared between three distant parties, namely Alice (A), Bob (B), and Charlie (C). We will show here that the W-class of state may be a good candidate for the secret sharing protocol even though it is not helpful in quantum teleportation protocol \cite{agarwal_2006, joo_2003}. Moreover, we will use a symmetric QCM to unmask the secret sent by the sender. The introduced protocol may be described in a few steps, which are given in the following subsections.
	\subsubsection{Construction of the composite system involving the single qubit in which the secret is encoded and the shared three-qubit state}
	Let us consider the secret sharing protocol between three parties: Alice (A), Bob (B), and Charlie (C). We assume that the sender, Alice, wishes to send the secret to the receiver, Charlie, via the mediator, Bob. 
	To execute the protocol, Alice encodes the secret information in a single-qubit state of the form
	\begin{eqnarray}
		\ket{\phi}_A=a\ket{0}_A+b\ket{1}_A,~~\nor{a}+\nor{b}=1 
		\label{alice_secret_new}
	\end{eqnarray}
	Alice's task is to send the secretly prepared state to Charlie when Bob helps Charlie in this regard. We may note here a crucial point is that Bob can help Charlie only when he gets some partial information from Alice.\\ 
	To start the protocol, Alice prepared the three-qubit W-class of states $\ket{W}_{ABC}$, which is of the following form
	\begin{eqnarray}
		\ket{W}_{ABC}= \alpha \ket{001}_{ABC} +\beta \ket{010}_{ABC}+ \gamma \ket{100}_{ABC}
		\label{wclass}
	\end{eqnarray}
	where the normalization condition gives $\nor{\alpha}+\nor{\beta}+\nor{\gamma}=1$. Alice keeps the first qubit to herself and distributes the second and third qubit to Bob and Charlie, respectively. We assume that the state parameters $\alpha,\beta$ and $\gamma$ are known to Alice and Charlie only. The reason for this would be clear in the later section.\\
	Since Alice wishes to communicate the secret $\ket{\phi}_A$ to Charlie via Bob using the shared W-class of state $\ket{W}_{ABC}$, so she constructs a composite system $\ket{\psi}_{AABC}$ by attaching the secret state $\ket{\phi}_A$ to the shared state $\ket{W}_{ABC}$. The four-qubit composite system is given by
	\begin{equation}
			\eqalign{
				\ket{\psi}_{AABC}=\ket{\phi}_A \otimes \ket{W}_{ABC} \cr
				= a\alpha \ket{0001}_{AABC}+a\alpha \ket{0010}_{AABC} +a\beta \ket{0100}_{AABC}\cr
				+b\alpha \ket{1001}_{AABC}+ b\alpha \ket{1010}_{AABC}+b\beta \ket{1100}_{AABC}\cr
				=\frac{1}{\sqrt{2}}\bigg(\ket{\phi^+}[\ket{0}_B (a\alpha\ket{1}_C+ b\gamma\ket{0}_C)+a\beta \ket{1}_B \ket{0}_C]\cr
				+\ket{\phi^-}[\ket{0}_B (a\alpha\ket{1}_C- b\gamma\ket{0}_C)+a\beta \ket{1}_B \ket{0}_C]\cr
				+\ket{\psi^+}[\ket{0}_B (a\gamma\ket{0}_C+b\alpha\ket{1}_C)+b\beta\ket{1}_B \ket{0}_C]\cr
				+\ket{\psi^-}[\ket{0}_B (a\gamma\ket{0}_C- b\alpha\ket{1}_C)-b\beta \ket{1}_B \ket{0}_C]\bigg)}
		\label{composite_new}	
	\end{equation}
	where $\ket{\phi^{\pm}}=\frac{1}{\sqrt{2}}(\ket{00}\pm \ket{11})$ and $\ket{\psi^{\pm}}=\frac{1}{\sqrt{2}}(\ket{01}-\ket{10})$ denotes the two-qubit Bell states.
	\subsubsection{Bell state measurement performed by Alice}
	
	In the composite system $\ket{\psi}_{AABC}$, the first two qubits are possessed by Alice, and the third and fourth qubits belong to Bob and Charlie, respectively. \\
	Alice then performs a two-qubit Bell state measurement on her two qubits of the composite system $\ket{\psi}_{AABC}$ and as a consequence, Bob and Charlie share a two-qubit entangled state among themselves as per Alice's Bell state measurement result. It may be easy to find out that at this stage, it is not possible either for Bob or for Charlie to retrieve Alice's secret alone. Therefore, they need to cooperate with each other to get the secret. Once they agree to cooperate with each other, Alice chooses one of the two parties (either Bob or Charlie) to share her measurement result. Let us assume that Alice chooses Bob to send the Bell state measurement outcome. Alice then sends two classical bits only to Bob to inform him about her Bell state measurement outcome. According to the Bell state measurement outcome, the two-qubit entangled state shared between Bob and Charlie is given in the table below (Table 1).\\   
	\begin{table*}[ht]
		\centering
		\resizebox{0.75\textwidth}{!}{%
			\renewcommand{\arraystretch}{1.5}
			\begin{tabular}{|c|c|c|}
				\hline
				\begin{tabular}[c]{@{}c@{}}Alice's Bell state \\ measurement outcome\end{tabular} & \begin{tabular}[c]{@{}c@{}}Two-qubit entangled state \\ between Bob and Charlie\end{tabular} & \begin{tabular}[c]{@{}c@{}}Alice communicates with \\ Bob by sending two C-bits\end{tabular} \\ \hline
				$\ket{\phi^+}_{AA}$ & $a\alpha\ket{01}_{BC}+a\beta\ket{10}_{BC}+b\gamma\ket{00}_{BC}$ & $0_A0_A$ \\
				\hline
				$\ket{\phi^-}_{AA}$ & $a\alpha\ket{01}_{BC}+a\beta\ket{10}_{BC}-b\gamma\ket{00}_{BC}$ & $1_A1_A$ \\
				\hline
				$\ket{\psi^+}_{AA}$ & $a\gamma\ket{00}_{BC}+b\alpha\ket{01}_{BC}+b\beta\ket{10}_{BC}$ & $0_A1_A$ \\
				\hline
				$\ket{\psi^-}_{AA}$ & $a\gamma\ket{00}_{BC}-b\alpha\ket{01}_{BC}-b\beta\ket{10}_{BC}$ & $1_A0_A$
				\\
				\hline
			\end{tabular}%
		}
		\caption{Alice performs a two-qubit Bell state measurement on the composite system (\ref{composite_new}). The measurement outcome is described in column 1. As a result of the measurement, Bob and Charlie share a two-qubit state given in column 2. Alice also sends two classical bits (C-bits) to Bob, describing her measurement outcome.} 
		\label{table_alice measurement result_new}
	\end{table*}

	\subsubsection{Single qubit measurement performed by Bob}
	Depending upon Alice's two-qubit measurement result as $\ket{\phi^+},\ket{\phi^-},\ket{\psi^+}$ and $\ket{\psi^-}$, the shared state between Bob and Charlie can be re-expressed as
	\begin{eqnarray}
		\ket{\psi_{1}}_{BC}= \ket{0}_B (a\alpha\ket{1}_C+ b\gamma\ket{0}_C)+a\beta \ket{1}_B \ket{0}_C \label{bc_shared_phip_new}\\
		\ket{\psi_{2}}_{BC}= \ket{0}_B (a\alpha\ket{1}_C- b\gamma\ket{0}_C)+a\beta \ket{1}_B \ket{0}_C\label{bc_shared_phim_new}\\
		\ket{\psi_{3}}_{BC}= \ket{0}_B (a\gamma\ket{0}_C+b\alpha\ket{1}_C)+b\beta\ket{1}_B \ket{0}_C\label{bc_shared_psip_new}\\
		\ket{\psi_{4}}_{BC}= \ket{0}_B (a\gamma\ket{0}_C- b\alpha\ket{1}_C)-b\beta \ket{1}_B \ket{0}_C\label{bc_shared_psim_new}
	\end{eqnarray}
	Since Alice chooses Bob to share her Bell state measurement outcome so Bob now performs a single-qubit measurement on any one of these states (\ref{bc_shared_phip_new}-\ref{bc_shared_psim_new}) in the computational basis $\{\ket{0},\ket{1}\}$. 
	\begin{table*}[ht]
		\renewcommand{\arraystretch}{1.5}
			\begin{tabular}{|c|c|c|}
				\hline 
				\begin{tabular}[c]{@{}c@{}}Shared state between \\ Bob and Charlie\end{tabular} & \begin{tabular}[c]{@{}c@{}}Bob's measurement\\ outcome\end{tabular} &  \begin{tabular}[c]{@{}c@{}}State collapsed at \\ Charlie's end\end{tabular} \\
				\hline \hline
				\multirow{2}{*}{$\ket{0}_B (a\alpha\ket{1}_C+ b\gamma\ket{0}_C)+a\beta \ket{1}_B \ket{0}_C$} & $\ket{0} $ & $a\alpha\ket{1}_C+ b\gamma\ket{0}_C$ \\ \cline{2-3}
				& $\ket{1}$ & $\ket{0}_C$  \\ 
				\hline \hline
				\multirow{2}{*}{$\ket{0}_B (a\alpha\ket{1}_C- b\gamma\ket{0}_C)+a\beta \ket{1}_B \ket{0}_C$} & $\ket{0}$ & $a\alpha\ket{1}_C- b\gamma\ket{0}_C$ \\ \cline{2-3}
				& $\ket{1}$ & $\ket{0}_C$ \\
				\hline \hline
				\multirow{2}{*}{$\ket{0}_B (a\gamma\ket{0}_C+b\alpha\ket{1}_C)+b\beta\ket{1}_B \ket{0}_C$} & $\ket{0}$ & $a\gamma\ket{0}_C+b\alpha\ket{1}_C$ \\ \cline{2-3}
				& $\ket{1}$ & $\ket{0}_C$ \\
				\hline \hline
				\multirow{2}{*}{$\ket{0}_B (a\gamma\ket{0}_C- b\alpha\ket{1}_C)-b\beta \ket{1}_B \ket{0}_C$} & $\ket{0}$ & $a\gamma\ket{0}_C- b\alpha\ket{1}_C$ \\ \cline{2-3}
				& $\ket{1}$ & $\ket{0}_C$\\
				\hline 
			\end{tabular}%
		\caption{Bob performs single-qubit measurement on the shared two-qubit state obtained as a result of Alice's measurement in computational basis. The outcomes of Bob's measurement and the state collapsed at Charlie's end are given in columns 2 and 3, respectively.}
		\label{table_bob and charlie shared_new}
	\end{table*}
	
	Due to the single qubit measurement, the two cases might arise.\\
	\textbf{Case-I:} If Bob obtains his measurement outcome as $\ket{1}$, then the protocol can be aborted, and it has to restart again.\\
	\textbf{Case-II:} If Bob obtains the measurement outcome as $\ket{0}$, then the protocol may succeed and proceeded in the following way. When Bob gets the single qubit measurement outcome $\ket{0}$, corresponding to Alice's Bell state measurement outcome, then any of the four single qubit states given in (\ref{bc_shared_phip_new}-\ref{bc_shared_psim_new}), collapses at Charlie's site. Bob then informs Charlie about the possible collapsed single qubit state by sending two classical bits $\{00,01,10,11\}$. At this stage also, one may find that Charlie is unable to retrieve the secret sent by Alice unless $\alpha=\beta=\gamma=\frac{1}{\sqrt{3}}$. The single qubit measurement performed by Bob and the state collapsed at Charlie's end has been summarized in Table-II.     \\

	\subsubsection{Charlie retrieves the secret using a QCM and performing Bell state measurement}
	In the previous step, if $\alpha \neq \gamma$, then we will show that Charlie can still retrieve the secret by using a QCM.\\
	Thus, Charlie makes use of a QCM for secret extraction from the single-qubit state that he possesses. Since the shared state parameters $\alpha$ and $\gamma$ are known to Charlie, so he uses them to design the QCM transformation  (\ref{qcm_gen}) in the following way
	\begin{equation}
		\begin{split}		
			\ket{0}\ket{0}\ket{Q}\longrightarrow \big(\ket{00}+\alpha(\ket{01}+\ket{10})\big)\ket{Q_0}\\		\ket{1}\ket{0}\ket{Q}\longrightarrow \big(\ket{11}+\gamma(\ket{01}+\ket{10})\big)\ket{Q_1}
		\end{split}
		\label{qcm_new}
	\end{equation} 
	The unitarity of the transformation (\ref{qcm_new}) gives
	\begin{equation}
		\begin{split}
			\braket{Q_0}{Q_0}=\frac{1}{1+2\nor{\alpha}} \\
			\braket{Q_1}{Q_1}=\frac{1}{1+2\nor{\gamma}} 
			\label{unitarity_qcm_new}
		\end{split}
	\end{equation}
	To move ahead of the protocol, let us suppose that Bob gets the measurement outcome $\ket{0}$ when he performs a single qubit measurement on the state $\ket{\psi_{1}}$ given in (\ref{bc_shared_phip_new}). As a result, Charlie receives the state $a\alpha\ket{1}+b\gamma\ket{0}$ and then applies the QCM defined in (\ref{qcm_new}) on the received state. He then obtains the following transformed state
	\begin{equation}
		\begin{split}
			(a\alpha\ket{1}+b\gamma\ket{0})\ket{0}\ket{Q} \longrightarrow &\frac{1}{\sqrt{2}}\ket{\phi^+}(a\alpha \ket{Q_1}+b\gamma\ket{Q_0})\\
			&+\frac{1}{\sqrt{2}}\ket{\phi^-}(b\gamma\ket{Q_0}-a\alpha\ket{Q_1})\\
			&+\sqrt{2}\alpha\gamma\ket{\psi^+} (a\ket{Q_1}+b\ket{Q_0})
		\end{split}
	\end{equation}
Charlie then perform measurement in the Bell basis and perform single qubit unitary transformation i.e. pauli operation to retrieve the secret, which was sent by Alice. The following scheme is used by Charlie

\begin{table}[H]
	\centering
		\begin{tabular}{|c|c|} 
			\hline
			\begin{tabular}[c]{@{}c@{}}Classical bits sent \\by Bob\end{tabular} & \begin{tabular}[c]{@{}c@{}}Applied single qubit\\unitary operation\end{tabular}  \\ 
			\hline
			00    &  $\sigma_x$ \\ 
			\hline
			11 &  $\sigma_z \sigma_x$    \\ 
			\hline
			01 & $I$     \\ 
			\hline
			10 & $\sigma_z$    \\
			\hline
	\end{tabular}
	\caption{Decoding scheme used by Charlie based on two classical bits he received from Bob.}
\end{table}

In a similar fashion, if Charlie received the state either in the form of $a\alpha\ket{1}_C- b\gamma\ket{0}_C$ or $a\gamma\ket{0}_C+ b\alpha\ket{1}_C$ or $a\gamma\ket{0}_C- b\alpha\ket{1}_C$ after Bob's measurement then the unitary operation that has to be applied on the cloned state is given in the Table-IV. \\
					\begin{table*}[ht]
						\renewcommand{\arraystretch}{2}
						\resizebox{\textwidth}{!}{%
							\begin{tabular}{||c|c|c|c||}
								\hline
								\hline
								\begin{tabular}[c]{@{}c@{}}State obtained by Charlie after \\ Bob's single-qubit measurement\end{tabular} & State obtained by Charlie after applying QCM & \begin{tabular}[c]{@{}c@{}}The state obtained if the Bell state \\ measurement outcome is $\ket{\psi^+}$\end{tabular} & \begin{tabular}[c]{@{}c@{}}Unitary operation applied on the \\ state obtained depending upon the\\  two classical bits send by Bob\end{tabular} \\
								\hline
								\hline
								\multirow{2}{*}{$a\alpha\ket{1}_C+b\gamma\ket{0}_C$} & $\frac{1}{\sqrt{2}}\ket{\phi^+}(a\alpha \ket{Q_1}+b\gamma\ket{Q_0})+\frac{1}{\sqrt{2}}\ket{\phi^-}(b\gamma\ket{Q_0}-a\alpha\ket{Q_1})$ & \multirow{2}{*}{$a\ket{Q_1}+b\ket{Q_0}$} & \multirow{2}{*}{$\sigma_z$} \\
								& $+\sqrt{2}\ket{\psi^+}\alpha\gamma (a\ket{Q_1}+b\ket{Q_0})$ &  &  \\
								\hline
								\multirow{2}{*}{$a\alpha\ket{1}_C- b\gamma\ket{0}_C$} & $\frac{1}{\sqrt{2}}\ket{\phi^+}(a\alpha \ket{Q_1}-b\gamma\ket{Q_0})+\frac{1}{\sqrt{2}}\ket{\phi^-}(-b\gamma\ket{Q_0}-a\alpha\ket{Q_1})$ & \multirow{2}{*}{$a\ket{Q_1}-b\ket{Q_0}$} & \multirow{2}{*}{$\sigma_z\sigma_x$} \\
								& $+\sqrt{2}\ket{\psi^+}\alpha\gamma (a\ket{Q_1}-b\ket{Q_0})$ &  &  \\
								\hline
								\multirow{2}{*}{$a\gamma\ket{0}_C+ b\alpha\ket{1}_C$} & $\frac{1}{\sqrt{2}}\ket{\phi^+}(a\gamma \ket{Q_0}+b\alpha\ket{Q_1})+\frac{1}{\sqrt{2}}\ket{\phi^-}(a\gamma\ket{Q_0}-b\alpha\ket{Q_1})$ & \multirow{2}{*}{$a\ket{Q_0}+b\ket{Q_1}$} & \multirow{2}{*}{$I$} \\
								& $+\sqrt{2}\ket{\psi^+}\alpha\gamma (a\ket{Q_0}+b\ket{Q_1})$ &  &  \\
								\hline
								\multirow{2}{*}{$a\gamma\ket{0}_C- b\alpha\ket{1}_C$} & $\frac{1}{\sqrt{2}}\ket{\phi^+}(a\gamma \ket{Q_0}-b\alpha\ket{Q_1})+\frac{1}{\sqrt{2}}\ket{\phi^-}(a\gamma\ket{Q_0}+b\alpha\ket{Q_1})$ & \multirow{2}{*}{$a\ket{Q_0}-b\ket{Q_1}$} & \multirow{2}{*}{$\sigma_x$} \\
								& $+\sqrt{2}\ket{\psi^+}\alpha\gamma (a\ket{Q_0}-b\ket{Q_1})$ &  & \\
								\hline \hline
							\end{tabular}%
						}
						\caption{Charlie applies a QCM defined in (\ref{qcm_new}) on the single-qubit state he has. He performs a two-qubit Bell state measurement on the state obtained in column 2, and if his measurement outcome is $\ket{\psi^+}$, he applies a unitary transformation according to the classical bits received from Bob.}
						\label{table_charlie qcm_new}
					\end{table*}
					
Equation (5) and equation (24) are basically representing the same equations with $p=\alpha$ and $q=\gamma$. Therefore, the assumption made in equation (5) is also valid in equation (24). Thus, we can identify the two machine state vectors $\ket{Q_0}$ and $\ket{Q_1}$ at the output of the cloning transformation as $\ket{Q_0}\equiv \ket{0}$ and $\ket{Q_1} \equiv \ket{1}$.

					\subsection{Analysis of the proposed controlled QSDC protocol}
				In the previous section, we find some cases where we have to abort the controlled QSDC protocol in the midway. Thus, there is a possibility of failure of the proposed controlled QSDC protocol. Hence, we would like to analyze the probability of success of the proposed controlled QSDC protocol and also would like to investigate how the quality of the copying machine is related to the probability of success.
					
					\subsubsection{Probability of success of the controlled QSDC scheme}
					In the proposed controlled QSDC protocol, we have studied that Bob and Charlie after agreeing to co-operate with each other can decode Alice's secret using a QCM constructed by Charlie. However, there are cases when either Bob or Charlie have to abort the protocol. Thus, it is enough to consider only the cases when the extraction of the secret is successful at Charlie's end. By considering only those case where the controlled QSDC scheme is successful, we get the probability of success, which is given by
					
					\begin{eqnarray}
						P_s=&\sum_{\alpha_A} \bigg(P[\text{Two-qubit measurement outcome} ~\alpha_A]\times\notag \\ 
						&P[\text{single-qubit measurement outcome} ~\ket{0}_B] \times \notag \\ 
						&P[\text{two-qubit measurement outcome} ~\ket{\psi^+}_C]\bigg)\notag \\
						=& 4\nor{\alpha}\nor{\gamma}
						\label{probsucc_new}
					\end{eqnarray}
					where $\alpha_A\in \{\ket{\phi^+}_A\bra{\phi^+},\ket{\phi^-}_A\bra{\phi^-},\ket{\psi^+}_A\bra{\psi^+},\ket{\phi^-}_A\bra{\psi^-}\}$
					
					\subsubsection{Linkage between the probability of success and the quality of the QCM}
					In this section, we will investigate if anyway the quality of the copy emerging out of the quantum copying machine effects the probability of success of the proposed controlled QSDC scheme. In other words, we can ask the following question: Does there exist any relationship between the probability of success denoted by $P_{s}$ and the average HS distance denoted by $\overline{D_a}$?\\
					Recalling (\ref{avhsdis}) with $\nor{p}=\nor{\alpha}$ and $\nor{q}=\nor{\gamma}$, $\overline{D_a}$ can be re-expressed as
					\begin{eqnarray}
						\overline{D_a}=
						\frac{4}{3}\bigg[\frac{|\gamma|^4}{(1+2\nor{\gamma})^2}+\frac{|\alpha|^4}{(1+2\nor{\alpha})^2}\bigg]+\frac{1}{3}
						\label{avhsdis1_new}
					\end{eqnarray}
					Using (\ref{probsucc_new}) and (\ref{avhsdis1_new}), we can re-express the average HS distance $\overline{D_a}$ in terms of the probability of success $P_{s}$ and we will denote it by $\overline{D_s}$ given as
					
					\begin{eqnarray}
						\overline{D_s}= \frac{4}{3}\left[\dfrac{\left(1-\nor{\beta}+\dfrac{P_s}{2}\right)^2+\dfrac{P_s^2}{4}-\dfrac{P_s}{2}}{(1+2\nor{\gamma})^2(1+2\nor{\alpha})^2}\right]+\frac{1}{3}
						\label{dabar_new}
					\end{eqnarray}
					
					In the previous subsection, we have noted that $\overline{D_a}\geq \frac{1}{3}$ for any cloning machine parameter $\nor{p}$ and $\nor{q}$. But since in the proposed controlled QSDC protocol, $\nor{p}$ and $\nor{q}$ have been replaced by the three-qubit sharing state parameter $\nor{\alpha}$ and $\nor{\gamma}$ so it would be interesting to find out the possible range of $\nor{\alpha},\nor{\beta},\nor{\gamma}$ and $P_s$ for which we can make $\overline{D_s}\leq \frac{1}{3}$. To do this, we need to check whether there exists any range of $\nor{\beta}$ and $P_{s}$ for which the quantity $(1-\nor{\beta}+\frac{P_s}{2})^2+\frac{P_s^2}{4}-\frac{P_s}{2}$ satisfies the inequality
					\begin{eqnarray}
						\left(1-\nor{\beta}+\frac{P_s}{2}\right)^2+\frac{P_s^2}{4}-\frac{P_s}{2}\leq 0
						\label{ineq_new}
					\end{eqnarray}
					Since the left hand side of the inequality (\ref{ineq_new}) is quadratic in $P_{s}$ so by solving it, we get

						\begin{eqnarray}
							-\left(\frac{1}{2}-\nor{\beta}\right)- \frac{1}{2}\sqrt{12\nor{\beta}-4|\beta|^4-7} \leq P_s \leq -\left(\frac{1}{2}-\nor{\beta}\right)+ \frac{1}{2}\sqrt{12\nor{\beta}-4|\beta|^4-7}
							\label{rangeps_new}
						\end{eqnarray}
					The above inequality (\ref{rangeps_new}) is valid only when 
					\begin{equation}
						\begin{split}
							\frac{3-\sqrt{2}}{2}\leq \nor{\beta} \leq 1
						\end{split}
						\label{mod_beta_range_new}
					\end{equation}
					When we vary $\nor{\beta}$ in the interval $[\frac{3-\sqrt{2}}{2},1]$, we find from (\ref{rangeps_new}) that the probability of success $P_s$ also varies within the interval $[0,1]$, as it depends on the value of $\nor{\beta}$.\\
					We are now in a position to find the ranges for $\nor{\alpha}$ and $\nor{\gamma}$ with the help of normalization condition $\nor{\alpha}+\nor{\gamma}+\nor{\beta}=1$ and using the lower bound of $\nor{\beta}$. Therefore, we get
					\begin{eqnarray}
						0\leq \nor{\alpha}+\nor{\gamma}\leq \frac{1}{\sqrt{2}}-\frac{1}{2}
						\label{mod_alpha_gamma_range_new}
					\end{eqnarray}
					Thus, if (\ref{mod_beta_range_new}) and (\ref{mod_alpha_gamma_range_new}) are satisfied simultaneously, the average HS distance between the original and the copy would be less than $\frac{1}{3}$ and hence making it more efficient than Wootters-Zurek QCM. The relationship between $\overline{D_s}$ and $P_s$ is shown in Fig.1.

					\begin{figure}
						\centering
						\includegraphics[width=1\linewidth]{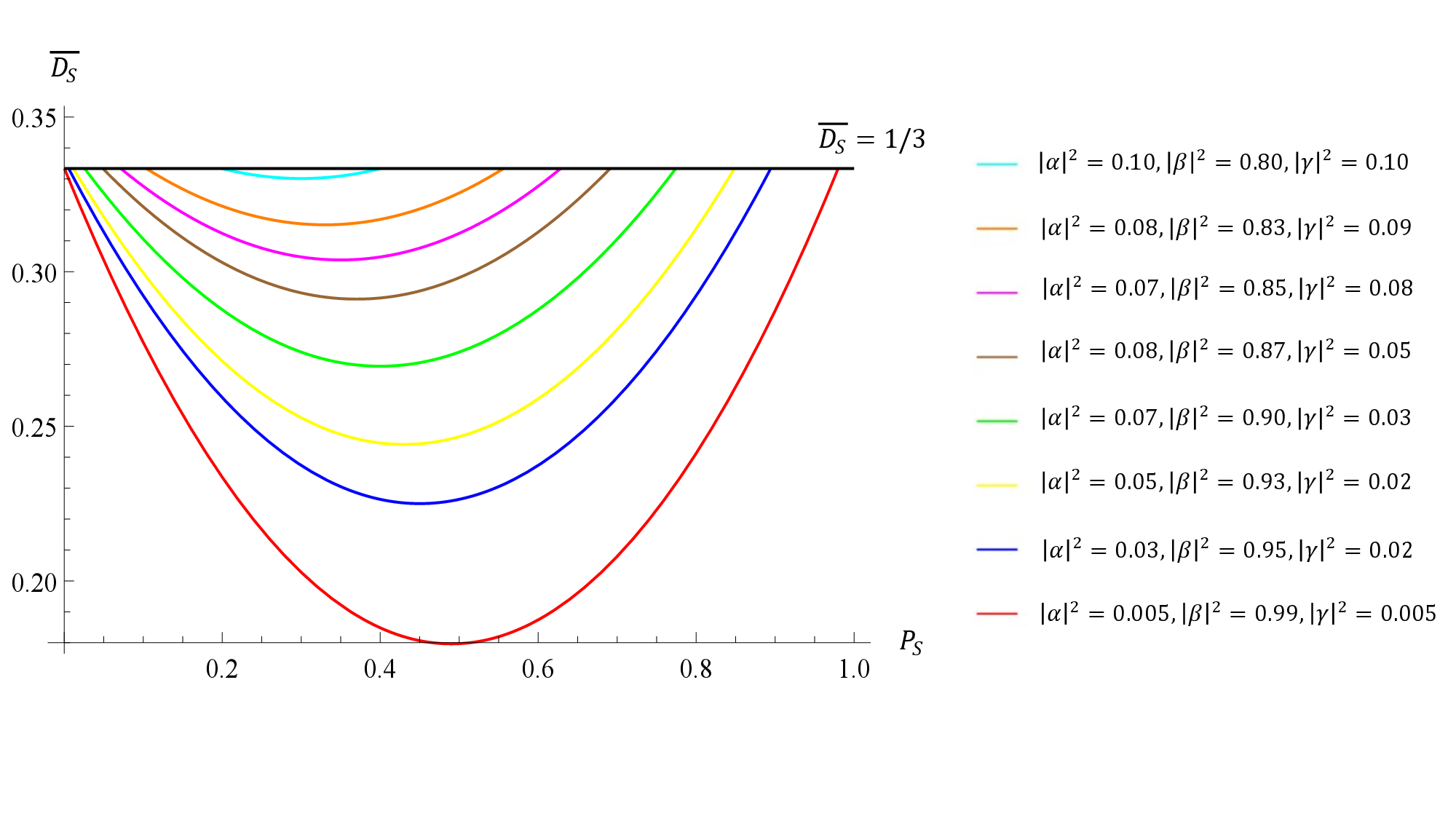}
						\caption{The graph between $\overline{D_s}$ and $P_s$  given by (\ref{dabar_new}) is plotted for different values of $\nor{\alpha},\nor{\beta}$ and $\nor{\gamma}$ for the ranges given in (\ref{mod_beta_range_new}) and (\ref{mod_alpha_gamma_range_new}). The straight line represents the value of $\overline{D_s}$ corresponding to the W-Z QCM.}
						\label{figure_ds bar_ps graph}
					\end{figure}
					
					\section{Relationship between concurrence fill and the probability of success of the controlled QSDC scheme}
					In this section, we would like to establish the connection between the amount of entanglement in the shared three-qubit W class of state quantified by concurrence fill and the probability of success of the proposed controlled QSDC scheme. We investigate whether the amount of entanglement in the shared state impacts the probability of success of the controlled QSDC scheme. In the proposed controlled QSDC scheme, we are using the W class of state to share the secret, and since tangle is unable to quantify the amount of entanglement in the W class of state, so we have to look for another measure that may quantify the amount of entanglement in a pure three-qubit W class of state.
					\subsection{Concurrence fill}
					Here, we will use an entanglement measure that may quantify the amount of entanglement in GHZ as well as the W class of state. To achieve this task, we use the \textit{concurrence fill}, a measure of entanglement, defined as follows \cite{xie_2021}
					\begin{eqnarray}
						F_{ABC}=\left[\frac{16}{3}Q(Q-C^2_{A(BC)})(Q-C^2_{B(AC)})(Q-C^2_{C(AB)}) \right] ^{1/4}
						\label{concfill}
					\end{eqnarray}
					where the quantities $Q, C_{A(BC)},C_{B(AC)}$ and $C_{C(AB)}$ are given by
					\begin{eqnarray}
						Q=\frac{1}{2}(C^2_{A(BC)}+C^2_{B(AC)}+C^2_{C(AB)})
						\label{q_entanglement_measure}
					\end{eqnarray}
					\begin{eqnarray}
						C_{i(jk)}=4 ~\text{det}(\rho_i), ~i,j,k=A,B,C;i\neq j\neq k    
					\end{eqnarray}
					The pre-factor $16/3$ ensures the normalization of the concurrence fill, i.e., $F_{ABC}\in [0,1]$. Now, we are in a position to quantify the amount of entanglement in a W class of state given in (\ref{wclass}) using concurrence fill.\\
					For a W class of state, we have
					\begin{eqnarray}
						C^2_{A(BC)}=& 4\nor{\gamma}-4|\gamma|^4\label{cabc2}\\
						C^2_{B(AC)}=& 4\nor{\beta}-4|\beta|^4\label{cbac2}\\
						C^2_{C(AB)}=& 4\nor{\alpha}-4|\alpha|^4\label{ccab2}
					\end{eqnarray}
					Using \eqref{cabc2}-\eqref{ccab2} and the normalization condition $\nor{\alpha}+\nor{\beta}+\nor{\gamma}=1$, we have
					\begin{eqnarray}
						Q=2(1-|\alpha|^4-|\beta|^4-|\gamma|^4)
					\end{eqnarray}
					The concurrence fill $F_{ABC}$ can be re-expressed in terms of $\nor{\alpha}$, $\nor{\beta}$ and $\nor{\gamma}$ as  
					
						\begin{equation}
							\begin{split}
								F_{ABC}=\frac{4}{\sqrt[4]{3}}\bigg(&(1-|\alpha|^4-|\beta|^4-|\gamma|^4)(1-|\beta|^4-|\alpha|^4+|\gamma|^4-2\nor{\gamma})\\
								&(1-|\alpha|^4-|\gamma|^4+|\beta|^4-2\nor{\beta})(1-|\gamma|^4-|\beta|^4+|\alpha|^4-2\nor{\alpha}) \bigg)^{1/4}
							\end{split}
						\end{equation}
					
					Let us now calculate the value of $F_{ABC}$ for some particular three-qubit state that belong to the W class of state.\\
					(i) When $\nor{\alpha}=\nor{\beta}=\nor{\gamma}=\frac{1}{3}$, therefore the value of $F_{ABC}$ for W state is given as 
					\begin{eqnarray}
						F_{ABC}= 0.88889
						\label{fabc_general_state}
					\end{eqnarray}
					(ii) Let us consider another W class of state, which is defined as  \cite{agarwal_2006}
					\begin{eqnarray}
						\ket{W_n}=\frac{1}{\sqrt{2+2n}}(\ket{100}+\sqrt{n}e^{i\gamma}\ket{010}+\sqrt{n+1}e^{i\delta}\ket{001})
					\end{eqnarray}
					where $n=1,2,3,\ldots$. $\gamma$ and $\delta$ are denoting phases. Therefore, $F_{ABC}$ for the above-defined W class of state is given as
					\begin{eqnarray}
						F_{ABC}=2\left(\frac{n^2 (n^2+3n+1)}{3(1+n)^6}\right)^{1/4}
					\end{eqnarray}
					It can be observed from Fig. 2 that when the value of $n$ increases, the amount of entanglement in the state $\ket{W_n}$ decreases. 
					\begin{figure}[H]
							\centering
							\includegraphics[width=1\linewidth]{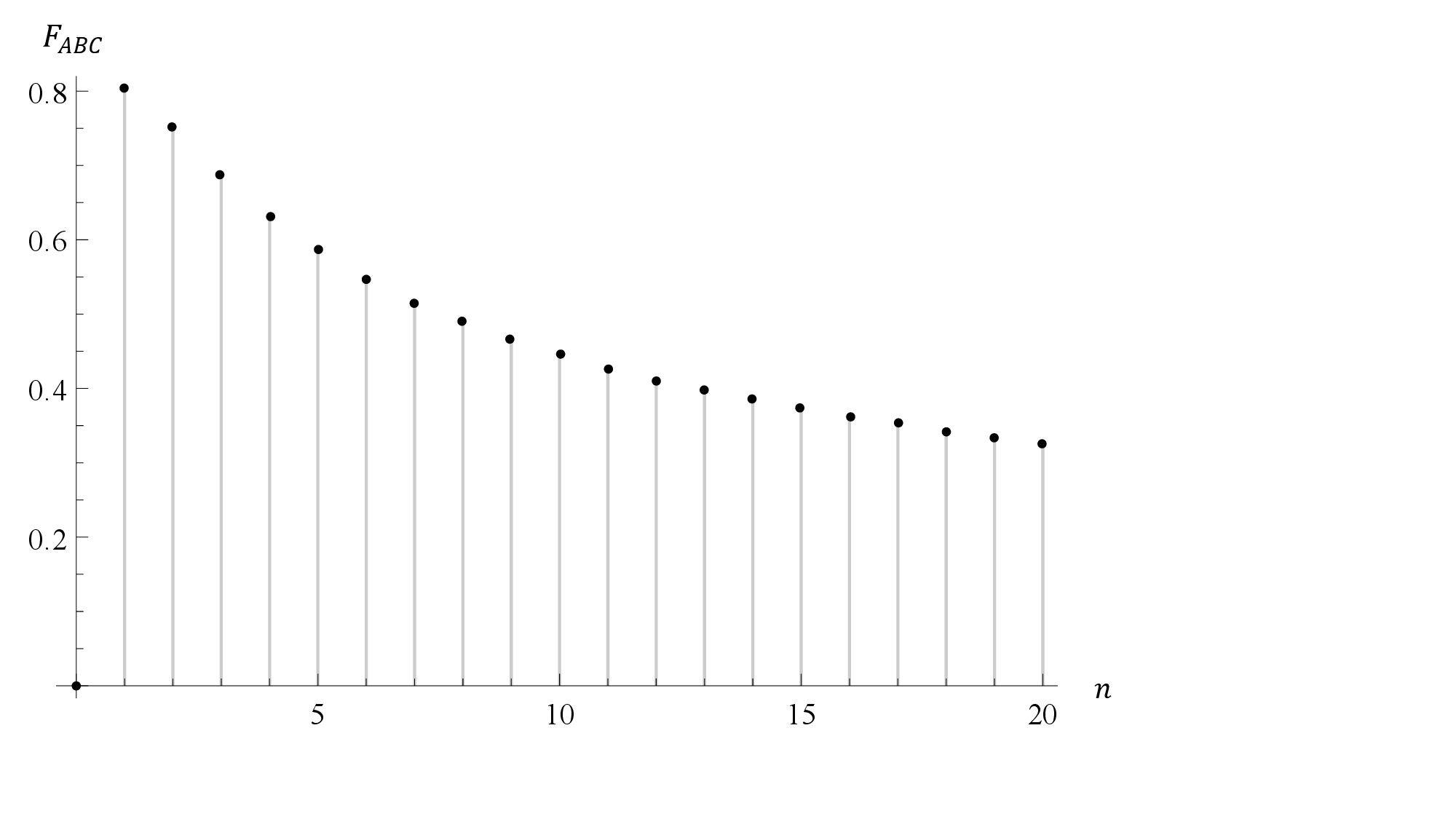}
							\caption{Graph between $F_{ABC}$ and $n$ is drawn. As the value of $n$ increases, the value of $F_{ABC}$ decreases.}
							\label{fig:enter-label}
						\end{figure}
					\subsection{Relationship between the concurrence fill and the probability of success of the controlled QSDC scheme}
					In this section, we show how the probability of success of the proposed controlled QSDC scheme varies with the amount of entanglement contained in the three-qubit W class of state shared between the three parties, Alice, Bob, and Charlie.\\
					To start with, let us first recall $P_{s}$, the probability of success of the controlled QSDC scheme, which is given in (\ref{probsucc_new}). Now, we are in a position to express the values of $C^2_{A(BC)}$, $C^2_{B(AC)}$ and $C^2_{C(AB)}$ in terms of $P_{s}$ and shared state parameters $\nor{\alpha}, \nor{\beta}$ and $\nor{\gamma}$, which are given by
					\begin{eqnarray}
						C^2_{A(BC)}=& P_s\left(\frac{1-\nor{\gamma}}{\nor{\alpha}}\right) \label{c2abc}\\
						C^2_{B(AC)}=& 4\nor{\beta}(1-|\beta|^2) \label{c2bac}\\
						C^2_{C(AB)}=& P_s\left(\frac{1-\nor{\alpha}}{\nor{\gamma}}\right) \label{c2cab}
					\end{eqnarray}
					Using \eqref{c2abc}-\eqref{c2cab}, the expression for $Q$ given in (\ref{q_entanglement_measure}) reduces to
					\begin{eqnarray}
						Q=P_s+4\nor{\beta}-4|\beta|^4
					\end{eqnarray}
					Therefore, the concurrence fill $F_{ABC}$ can then be re-expressed in terms of $P_{s}$ and $\nor{\beta}$ as  
					\begin{eqnarray}
						F_{ABC}=\left(\frac{64}{3}P_s^2|\beta|^4(P_s+4\nor{\beta}-4|\beta|^4)\right)^{1/4}
						\label{fabc1}
					\end{eqnarray}
					Since our interest lies in studying the effect of entanglement quantified by $F_{ABC}$ on the success probability $P_{s}$, we can re-write (\ref{fabc1}) as a cubic equation in $P_{s}$, which is given by
					\begin{eqnarray}
						P_{s}^{3}+4\nor{\beta}(1-|\beta|^2)P_{s}^{2}-\frac{3F_{ABC}^{4}}{64|\beta|^4}=0
						\label{psplusfacb}
					\end{eqnarray}
					Solving (\ref{psplusfacb}) using Cardon's method (given in Appendix-A), we get the solution as
					\begin{eqnarray}
						P_s=\frac{3F_{ABC}^4}{64 |\beta|^4}-\frac{128}{27}|\beta|^6(1-\nor{\beta})^3
						\label{ps_val_in_f}
					\end{eqnarray}
					with the following restriction, 
					\begin{eqnarray}
						F_{ABC}\geq \bigg(\frac{16384}{81}|\beta|^{10}(1-\nor{\beta})^3\bigg)^{1/4}
						\label{f_geq_restriction}
					\end{eqnarray}
					Since we know that the success probability lies between [0,1] so using (\ref{ps_val_in_f}) and (\ref{f_geq_restriction}), we obtain the lower and upper bound of $F_{ABC}$ as

						\begin{eqnarray}
							\bigg(\frac{16384}{81}|\beta|^{10}(1-\nor{\beta})^3\bigg)^\frac{1}{4} \leq F_{ABC}\leq \bigg(\frac{8192}{81}|\beta|^{10}(1-\nor{\beta})^3+\frac{64}{3}|\beta|^4\bigg)^\frac{1}{4}
							\label{f_inequality_for_ps}
						\end{eqnarray}
					
					Since the upper bounds of the concurrence fill $F_{ABC}$ for the W class of state has been achieved for the state $\ket{W}=\frac{1}{\sqrt{3}}(\ket{001}+\ket{010}+\ket{100})$ which is given by $0.88889$, so the upper bound of $F_{ABC}$ given in (\ref{f_inequality_for_ps}) must be less than or equal to 0.88889. This gives the range of $\nor{\beta}$ as $(0,0.17)$. For the given value of $\nor{\beta}$ lying in the interval $(0,0.17)$, we have drawn $P_s-F_{ABC}$ graph, which is given in Fig.3. The following facts can be observed from Fig.3, which is given below.\\
					\textbf{Observation-1:} For the value of $\nor{\beta}=0.017$, we observe that $P_s$ increases slowly but approaches to 1 for a very small range of $F_{ABC}$.\\
					\textbf{Observation-2:} For $\nor{\beta}=0.17$, we observe that $P_s$ increases rapidly from 0 to 1 for a greater range of $F_{ABC}$ as compared to the first case.
					
					\begin{figure}[H]
						\centering
						\includegraphics[width=1\linewidth]{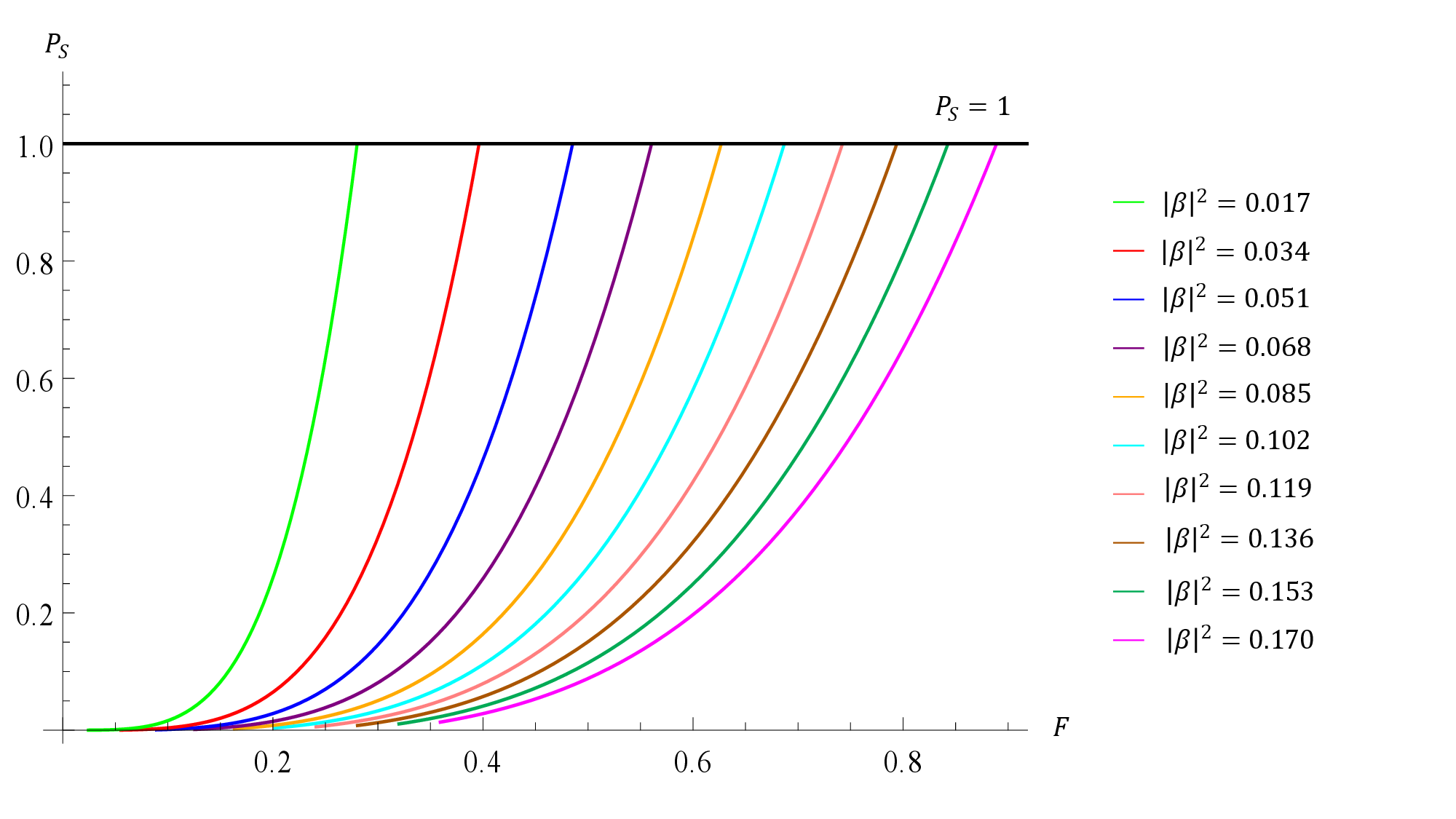}
						\caption{The $P_s-F_{ABC}$ graph is given with respect to the value of $\nor{\beta}$ lying in the interval $(0,0.17)$ and $F_{ABC}$ ranging according to the inequality (\ref{f_inequality_for_ps}) and less than equal to 0.88889.}
						\label{ps_f_figure}
					\end{figure}
					
				{\section{Discussion}
						In this section, we present the comparison of our study with the existing work. We also present the pros and cons of our study. Further, we provide a detailed discussion on various internal and external attacks that may be possible on the proposed QSDC scheme.
					\subsection{Comparison with the existing works}
						There are many works, for the development of QSDC scheme, available in the literature. We, therefore, compare our QSDC protocol with the others, which are given below:
                        
                        \begin{table}[H]
						\begin{tabular}{|p{0.5\linewidth}|p{0.5\linewidth}|}
							\hline
							\multicolumn{1}{|c|}{Existing Works} & \multicolumn{1}{c|}{Our Work} \\ \hline
							 Recently, in 2023, Liu et. al \cite{Liu_2023} proposed two QSDC protocols based on three-qubit W states that are developed under collective dephasing and rotation noise with the efficiency of $37.5\%$.  & Our proposed QSDC protocol uses a three-qubit W class of states as a quantum channel and we find, for a particular case, our protocol proves to be $66.6\%$ efficient.\\ \hline
						 Cao and Song \cite{cao_2006} proposed a QSDC protocol using four-qubit W state in which two users can communicate using Bell state measurements and classical communications. Later, in 2009, Dong et. al \cite{dong_2009} introduced the QSDC protocol using four qubit W class of states in a noisy channel.& Our protocol is better in the sense that it uses three-qubit W class of states instead of a four-qubit state, which are much more challenging to prepare. \\
                         \hline
							Various other works have been done towards QSDC protocols based on particular type of three-qubit state known as GHZ states \cite{gianni_2021}. & Since three-qubit W states are much more robust than the three-qubit GHZ states, we used W class of states in our proposed QSDC scheme.\\ \hline
                            & We have introduced the use of quantum cloning transformations in controlled QSDC protocols, which have not been included in existing QSDC protocols. We use quantum cloning transformation in our proposed QSDC protocol because it will help the receiver to extract the information sent by the sender.\\ \hline
						\end{tabular}
					\end{table}	

					\subsection{Pros and Cons of the proposed QSDC protocol}
						The pros and cons of our study are given below.}

					\begin{table}[H]
						\begin{tabular}{|p{0.5\linewidth}|p{0.5\linewidth}|}
							\hline
							\multicolumn{1}{|c|}{Pros} & \multicolumn{1}{c|}{Cons} \\ \hline
							The proposed QSDC protocol proves to be 66.6\% efficient for a particular choice of the parameters of the shared state between three parties, which is better than the earlier proposed QSDC scheme \cite{Liu_2023}. & In the proposed controlled QSDC protocol the receiver makes use of a quantum cloning machine (QCM) to retrieve Alice's secret. The QCM is described by an unitary transformation which require huge cost to implement it in the experiment.   \\ \hline
							Our QSDC protocol requires the use of minimum of three-qubit W class of states as a quantum channel. Three-qubit states are less challenging to prepare as compared to the four-qubit W class of states.& In a multi-party setting ($n>3$), as the number of receivers increases, the proposed QSDC protocol will become more complicated. This may decrease the efficiency of the proposed protocol. \\ \hline
							Since three-qubit W class of states are more robust than the three-qubit GHZ class of states, in terms of particle loss, so it can be advantageous to use them in the proposed controlled QSDC scheme. & For the execution of the proposed QSDC protocol, we require four classical bits that can be transmitted using classical technologies. Thus, it can be considered as a drawback of our QSDC protocol as there may exist other QSDC protocol which require fewer number of classical bits in comparison to ours. \\ \hline
						\end{tabular}
					\end{table}	
					
					\section{Security Analysis}
						
                        In the literature, there are two types of attacks, internal and external, on QSDC protocols. Generally, the inside attacker is more powerful than the outside attacker in terms of the availability of quantum resources of the protocol. Therefore, it can be concluded that if the protocol is safe from inside attackers, then it would be safer than outside attackers as well \cite{Xin_2024}. Thus, it would be interesting to discuss the various internal and external attacks on our proposed QSDC protocol. The proposed QSDC protocol is considered to be safe against any inside and outside attacks. 

                        \subsection{Inside attacks}
                        For the proposed controlled QSDC protocol, the insider attacks are of two types: attack by the receiver and attack by the controller, which are discussed below. 
                        
                        \subsubsection{Attack by the receiver:}
                        Let us assume that Charlie (receiver) is dishonest and tries to eavesdrop on the channel between Alice (sender) and Bob (controller). He somehow intercepts Bob's particle and performs a single-qubit measurement on it. By doing so, he tries to gain access to the controller's powers. Corresponding to the result obtained, Charlie obtains a single-qubit state and applies the prescribed QCM on it. There are four possible states that could collapse at Charlie's site (given in Table \ref{table_charlie qcm_new}). But, in order to retrieve Alice's secret, Charlie has to perform a single-qubit unitary operation on the collapsed state.\\
                        In spite of stealing Bob's qubit, Charlie won't be able to get any useful information because he is unaware of the two classical bits shared by the sender Alice to the controller Bob, which are necessarily required by Charlie to apply the unitary operation.\\
                        Therefore, our protocol can be considered as safe from the above discussed attack by the receiver.

                        \subsubsection{Attacks by the controller:}
                    Assume that Bob, the controller, is a dishonest participant and he does not share any information with the receiver Charlie about Alice's Bell state measurement results, i.e., the two classical bits. Then, Charlie won't be able to apply the correct unitary transformation in the last step of the protocol. But, this doesn't help Bob. The shared state parameters are only known to Alice and Charlie, and since Bob is unaware of the shared state parameters required for the construction of a quantum cloning machine, which restricts him from stealing the information.\\
                    Therefore, the proposed protocol is safe from this type of attack by the controller.
                    
                     \subsection{Outside Attacks}
				Let us assume that there is an outside attacker, say Eve. If Eve is present in between Bob and Charlie, she cannot achieve any useful information because there is no transmission of the qubit which carries secret message. Also, since the shared state parameters are only known to Alice and Charlie, she cannot clone Charlie’s particle even if she intercepts it. \\
                In summary, the proposed QSDC protocol is safe from both internal and external security damages and provides secure communication.\\ 
					
					\section{Conclusion}	
					To summarize, we have proposed a new controlled QSDC scheme between three parties, namely Alice, Bob, and Charlie, with the help of the W class of state and symmetric quantum cloning machine (QCM). In the proposed scheme, Alice wishes to send the secret information to Charlie via the controller Bob, who has partial information of the secret, i.e., to retrieve Alice's secret, Bob has to cooperate with Charlie. We have assumed that the shared state parameters are known only to Alice and Charlie in the proposed controlled QSDC scheme. On the other hand, if the shared state parameters are known to all three parties and in case if Charlie's qubit goes to Bob's place, then in this situation, Bob (if he is dishonest) can take advantage of it. He (Bob) may be able to construct a QCM and then perform all the steps of the proposed controlled QSDC scheme to find out Alice's secret without revealing the fact to Alice that Charlie is not present in the scheme. Therefore, it is necessary from the security point of view that only Alice and Charlie should know about the state parameters. Alice sends the Bell state measurement outcome to Bob and in this way, sharing partial information about the secret with Bob. On the other hand, the shared state parameters are only known to Alice and Charlie, which helps Charlie to prepare a symmetric QCM. Therefore, Bob and Charlie hold partial information related to Alice’s secret encoded in a quantum state. Thus, when Bob and Charlie agree to cooperate, then only Charlie can obtain Alice’s secret. Without getting cooperation from Bob, Charlie could not have extracted the information. Thus, Bob acting as the controller in the protocol. Being the controller, he cannot obtain any quantum information or secret message from the decoding process.\\
					The quality of the QCM was analyzed by studying the HS distance $D_a$. When an unknown input state is fed into the introduced QCM, the average HS distance $\overline{D_a}$ is found to be greater than or equal to $1/3$. Interestingly, if the parameters of the cloning machine tend towards 0, then the constructed QCM tends to the W-Z QCM. Further, we found that the proposed QCM may not be efficient for cloning in the sense that the average HS distance is not very small, but it plays a vital role in the introduced controlled QSDC scheme. Moreover, we observed that the probability of success of the proposed controlled QSDC scheme depends upon the parameters of the QCM. The ranges for success probability and state parameters were derived for which the QCM works better in the controlled QSDC scheme. In this regard, we have shown that there exist the values of the state parameters lying in the obtained ranges, for which $\overline{D_s}\leq 1/3$, making the constructed QCM a suitable choice for copying the input state for the proposed controlled QSDC protocol. Another interesting fact is that if we consider the shared state between the three parties of the form $\ket{\psi}=\alpha\ket{001}+\alpha\ket{010}+\beta\ket{100}$ with $2\nor{\alpha}+\nor{\beta}=1$, then we find that the shared state $\ket{\psi}$ will be useful for secret sharing but the HS distance $\overline{D_s}\geq 1/3$. Thus, in this case, the constructed QCM is not useful in copying the input state, but it can be used in the proposed controlled QSDC scheme to retrieve the secret.\\
					Moreover, we have studied the concurrence fill $F_{ABC}$ that can be used as a measure of three-qubit entanglement for the shared W-class of state. We further derived the relationship between the probability of success ($P_{s}$) of the introduced controlled QSDC scheme and $F_{ABC}$. We found that as the value of $\nor{\beta}$ increases, the value of $F_{ABC}$ also increases from 0 to $0.88889$. But in the perspective of $P_s$, we observed that as $\nor{\beta}$ varies from 0.017 to 0.17, $P_s$ approaches towards unity very slowly for a very small range of $F_{ABC}$, and then for a wider range of $F_{ABC}$, we found that $P_s$ increases rapidly from 0 to 1.
					
					\section{Data availability statement}
					Data sharing not applicable to this article as no datasets were generated or analysed during the current study.
					
					
                    \balance

					\section{Appendix}
					\begin{appendices}
						\section{Solution of cubic equation $P_s^3+P_s^2C_{B(AC)}^2-\frac{3F_{ABC}^4}{64 |\beta|^4}=0$}
						In this section, we discuss the Cardon's method used for finding the roots of the cubic equation in $P_s$, which is given in (\ref{psplusfacb}). The roots so obtained can be used to determine the relationship between the probability of success $(P_{s})$ of the controlled QSDC scheme and the amount of entanglement quantified by $F_{ABC}$ in the shared W class of state.  \\
						Using (\ref{c2bac}), we can re-express (\ref{psplusfacb}) as
						\begin{eqnarray}
							P_s^3+P_s^2C_{B(AC)}^2-\frac{3F_{ABC}^4}{64 |\beta|^4}=0
							\label{ps_val_in_f_a1}
						\end{eqnarray}
						Cardon's method can be used to find the roots of the general cubic equation of the form
						\begin{eqnarray}
							a_0 x^3+3a_1x^2+3a_2x+a_3=0
							\label{a0cubic_a1}
						\end{eqnarray}
						where $a_i~ (i=0,1,2,3)$ denotes the real numbers.\\
						Comparing (\ref{a0cubic_a1}) with (\ref{ps_val_in_f_a1}), we get
						\begin{equation}
							\begin{split}
								a_0=1, ~~~~~ a_1&=\frac{C_{B(AC)}^2}{3}\\
								a_2=0, ~~~~~ a_3&=-\frac{3F_{ABC}^4}{64|\beta|^4}
								\label{a_i_val_a1}
							\end{split}
						\end{equation}
						The cubic equation (\ref{a0cubic_a1}) can be re-expressed in the form
						\begin{eqnarray}
							z^3+3Hz+G=0
							\label{zcube_eq_a1}
						\end{eqnarray}
						where $H=a_0a_2-a_1^2$ and $G=a_0^2a_3-3a_0a_1a_2+2a_1^3$. Substituting the values of $a_0,a_1,a_2$ and $a_3$ for $H$ and $G$, we obtain, 
						\begin{equation}
							\begin{split}
								H&=\frac{-C_{B(AC)}^4}{9}\\
								G&=\frac{2}{27}C_{B(AC)}^6-\frac{3F_{ABC}^4}{64|\beta|^4}
								\label{h_g_val_a1}
							\end{split}
						\end{equation}
						Let us assume $P_s=u+v$. Cubing both sides, we get
						\begin{eqnarray}
							P_s^3-3uvP_s-(u^3+v^3)=0
							\label{ps_u_v_a1}
						\end{eqnarray}
						Upon comparing (\ref{zcube_eq_a1}) with (\ref{ps_u_v_a1}), we have $u^3v^3=-H^3$ and $u^3+v^3=-G$. Considering $u^3$ and $v^3$ as the roots of the quadratic equation, which is given by
						\begin{eqnarray}
							t^2-(u^3+v^3)t+u^3v^3=0
							\label{eq_t_a1}
						\end{eqnarray}
						The discriminant $D$ of the quadratic equation (\ref{eq_t_a1}) is given by 
						\begin{eqnarray}
							D=\dfrac{9}{4096}\dfrac{F_{ABC}^8}{|\beta|^8}-\dfrac{12}{1728}\dfrac{F_{ABC}^4}{|\beta|^4}C_{B(AC)}^6
							\label{det_a1}
						\end{eqnarray}
						Since the roots of the quadratic equation (\ref{eq_t_a1}) are real numbers lying within the interval $[0,1]$, so we must have $D\geq 0$. This gives
						\begin{eqnarray}
							F_{ABC}\geq \frac{16384}{81} |\beta|^{10}(1-\nor{\beta})^3
							\label{f_geq_ineq_a1}
						\end{eqnarray}
						Therefore, the roots of the quadratic equation (\ref{eq_t_a1}) are given by
						
							\begin{eqnarray}
								t=\frac{\dfrac{3F_{ABC}^4}{64|\beta|^4}-\dfrac{2}{27}C_{B(AC)}^6\pm \sqrt{\dfrac{9}{4096}\dfrac{F_{ABC}^8}{|\beta|^8}-\dfrac{12}{1728}\dfrac{F_{ABC}^4}{|\beta|^4}C_{B(AC)}^6}}{2}
							\end{eqnarray}
					
						The two roots of (\ref{eq_t_a1}) are $u^3$ and $v^3$, which can be given as 
						\begin{equation}
							\begin{split}
								u^3=\frac{\frac{3F_{ABC}^4}{64|\beta|^4}-\frac{2}{27}C_{B(AC)}^6+ \sqrt{\frac{9}{4096}\frac{F_{ABC}^8}{|\beta|^8}-\frac{12}{1728}\frac{F_{ABC}^4}{|\beta|^4}C_{B(AC)}^6}}{2}\\
								v^3=\frac{\frac{3F_{ABC}^4}{64|\beta|^4}-\frac{2}{27}C_{B(AC)}^6-\sqrt{\frac{9}{4096}\frac{F_{ABC}^8}{|\beta|^8}-\frac{12}{1728}\frac{F_{ABC}^4}{|\beta|^4}C_{B(AC)}^6}}{2}
							\end{split}
						\end{equation}
						
						The three cube roots of $u^3$ are $u,u\omega,u\omega^2$ and those of $v^3$ are $v,v\omega,v\omega^2$, where $\omega=(-1\pm i\sqrt{3})/2$ are the cube root of unity \cite{hari_2022_book}. Therefore, we get
						\begin{equation}
							\begin{split}
								u=\frac{\frac{3F_{ABC}^4}{64|\beta|^4}-\frac{2}{27}C_{B(AC)}^6+ \sqrt{\frac{9}{4096}\frac{F_{ABC}^8}{|\beta|^8}-\frac{12}{1728}\frac{F_{ABC}^4}{|\beta|^4}C_{B(AC)}^6}}{2}\\
								v=\frac{\frac{3F_{ABC}^4}{64|\beta|^4}-\frac{2}{27}C_{B(AC)}^6-\sqrt{\frac{9}{4096}\frac{F_{ABC}^8}{|\beta|^8}-\frac{12}{1728}\frac{F_{ABC}^4}{|\beta|^4}C_{B(AC)}^6}}{2}
								\label{u_v_val}
							\end{split}
						\end{equation}
						
						Finally, substituting the values of $u$, $v$ from (\ref{u_v_val}) and $C_{B(AC)}$ from (\ref{c2bac}) in $P_s=u+v$, we obtain
						\begin{equation}
							\begin{split}
								P_s= \frac{3}{64}\frac{F_{ABC}^4}{|\beta|^4}-\frac{128}{27}|\beta|^6(1-\nor{\beta})^3
							\end{split}
						\end{equation}
						
					\end{appendices}
					\end{document}